\documentclass[aps,eqsecnum,preprintnumbers]{revtex4}
\usepackage{axodraw}
\newcommand{\beq}{\begin{equation}}
\newcommand{\eeq}{\end{equation}}
\newcommand{\beqa}{\begin{eqnarray}}
\newcommand{\eeqa}{\end{eqnarray}}
\newcommand{\lslash}[1]{#1\llap/}

\newcommand{\Eq}[1]{Eq.\ (\ref{#1})}
\newcommand{\Eqs}[2]{Eqs.\ (\ref{#1}) and (\ref{#2})}
\newcommand{\Eqss}[3]{Eqs.\ (\ref{#1}), (\ref{#2}) and (\ref{#3})}
\newcommand{\Ref}[1]{Ref.\ \cite{#1}}
\newcommand{\Tr}{\mbox{Tr}\,}

\begin{document}

\preprint{hep-ph/0403121}
\title{
Nucleon contribution to the induced charge of neutrinos in a matter
background and a magnetic field
}
\author{Jos\'e F. Nieves}
\affiliation{Laboratory of Theoretical Physics,
Department of Physics, P.O. Box 23343,
University of Puerto Rico
R\'{\i}o Piedras, Puerto Rico 00931-3343}
\date{March 2004}

\begin{abstract}
We study the nucleon contribution to the electromagnetic
vertex function of neutrinos that propagate in a matter background
in the presence of a magnetic field. Starting from the one-loop
expression for the corresponding terms of the vertex function,
and taking into account the anomalous magnetic coupling of the nucleons,
we calculate the $B$-dependent part of the
form factors that determine the induced charge of the neutrino. 
A formula for the neutrino induced charge is obtained, and it
is evaluated for various illustrative situations. 
The terms due to the nucleons can be important in some cases,
depending on the physical conditions of the environment.
\end{abstract}

\maketitle

%
%
\section{Introduction and Summary}

In the physical contexts in which the effects of an electron background
are being considered, in principle a nucleon background is also involved.
In some situations, and depending on the effects being studied, 
the presence of the nucleons is not particularly relevant,
in which case they can be thought to form an inert background that
makes the medium electrically neutral overall, with no
further physical implications. For example, for a photon that
propagates through a matter background, the proton contribution
to the plasma frequency is insignificant compared to the electron one,
since the corresponding terms are proportional to the inverse of the
charged particle mass. In most cases, the effects of the nucleons
on the photon dispersion relation turn out to be unimportant.

The situation is different when we consider neutrinos instead
of photons. When a neutrino propagates in a background of
normal matter (electrons and nucleons), the contributions
of the various background particle species to the neutrino 
dispersion relation are comparable, each being proportional 
to the background particle number density \cite{raffeltbook}.
As is well known, the nucleon contribution is identical for the three neutrino
flavors, and it is not relevant in the context
of neutrino oscillations in matter that involve only the standard 
flavor neutrinos, since the oscillation phenomena depends on the difference
in the dispersion relations. However, for example,
if the so-called sterile neutrinos 
also take part in the oscillation phenomena, then the nucleon contribution
must be included
since it is of the same magnitude as the electron contribution.

Besides the dispersion relation, another quantity of interest is 
the effective electromagnetic coupling of the neutrino in the background 
medium \cite{oraevsky85,oraevsky87,semikoz87a,semikoz87b,np1,np2,sawyer}. 
That coupling is the basis of some of the
neutrino processes that can occur in a medium but not in the vacuum,
such as the plasmon decay and the Cherenkov radiation 
process \cite{raffeltbook}.
In addition, in the presence of a magnetic field,
the electromagnetic coupling induces a modification of 
the neutrino dispersion relation \cite{dnp1,kimetal,esposito,elmfors}
which is the basis for the asymmetric neutrino emission phenomena
that has been considered in various astrophysical 
contexts \cite{kicks1,kicks2,kicks3,kicks4,kicks5,semikoz,semikozE,nunokawa}.
The neutrino electromagnetic interactions also
modify the collective properties of a plasma, which can
lead to a variety of effects \cite{silvaetal,bento,nievesinstabilities,%
mohantynp,semikozalpha}.

For calculational purposes, the electromagnetic coupling
is parametrized by the form factors of the vertex 
function \cite{semikoz89,dn1,salati}. 
The calculations of the nucleon background contribution to the form factors
reveal that it can be of the same magnitude as the one due
to the background electrons \cite{dn2}. Thus, for example, in the context
of the asymmetric neutrino emission process already mentioned,
the nucleon contribution is unimportant if only the standard
flavor neutrinos are considered, but they are significant and must
be included if sterile neutrinos are involved.

There has been interest recently in the calculation of the
neutrino electromagnetic coupling in the presence of a magnetic field.
The calculation in Ref. \cite{raffelt} applies if the 
particle number densities in the medium are relatively small, such that they
can be neglected entirely. A calculation taking into account the
electrons in the background was performed in Ref. \cite{ganguly},
and it was extended and some aspects were clarified in Ref. \cite{nievesI}.
In Ref. \cite{nievesI} the focus was on the form
factors of the full vertex function, while Ref. \cite{ganguly} 
was focused on the particular term of the vertex function that 
can be identified as the induced neutrino charge. 
In spite of these differences, the calculations have in common
that they neglect the effects of the nucleons in the medium. For some
applications, such simplification may be appropriate. However,  
as we have mentioned above, that it is not always the case and, therefore,
the results of the pertinent calculations are needed. 
Such calculations are the focus of the present work.

Here we consider the effects of the nucleon background
on the electromagnetic coupling of the neutrino, in the presence
of a magnetic field $B$. In particular, we take into account the
anomalous magnetic moment interaction of the nucleons with the magnetic field
and, as in Refs. \cite{ganguly} and \cite{nievesI}, our goal is to determine
those corrections that are linear in $B$. We follow
the treatment of the latter reference, adapted to the present situation.
An essential ingredient in our previous calculation was
the linear approximation to the Schwinger formula for the
electron propagator in a magnetic field, extended to include
the presence of the electron background. Here we derive
the corresponding formula for the nucleon propagator, which
takes into account the anomalous magnetic coupling of the nucleons.
Using that propagator we obtain the expression for the
one-loop contribution to the neutrino vertex function. Although
the vertex function can be decomposed in terms of a set of tensors
as discussed in Ref. \cite{nievesI}, here we do not give explicitly
the integral formulas for all the associated form factors in
the general case, which are cumbersome and not illuminating.
For definiteness, we carry out in detail the evaluation of the
vertex function in the so-called \emph{static} and 
\emph{long wavelength limit}, which can be interpreted in terms of
the neutrino effective charge in the medium. We
give the formula for the nucleon contribution to that quantity,
in terms of momentum integrals over the nucleon number densities,
and in order to establish a point of comparison with the previous calculations
we evaluate it explicitly for various illustrative cases.
As we have anticipated,
the nucleon and the electron contributions can be comparable,
depending on the conditions of the environment.
Our calculations can be useful for the studies of the effects of the
neutrino electromagnetic interactions
in various astrophysical environments,
which have been considered in the literature.

%
%
\section{The linearized nucleon propagator}
\label{sec:propagator}

The propagator that was used in the calculation of
Ref. \cite{nievesI} was obtained by taking the Schwinger formula
for the electron propagator in a magnetic field, and expanding it
up to linear terms in $B$. We want to obtain the corresponding
formula for the nucleon propagator, and we follow closely the notation
of that reference. But in order to include
the anomalous magnetic moment coupling, here we proceed as follows.

In the presence of the magnetic field, the propagator of each
nucleon $f= n,p$ satisfies the equation
\beq
\label{coordeq}
\left[i\lslash{\partial} - e_f\lslash{A} - m_f - \frac{1}{2}\kappa_f\sigma
\cdot F\right]S(x,x^\prime) = \delta^{(4)}(x - x^\prime) \,,
\eeq
where $e_p = |e|$ and $e_n = 0$,
and $\kappa_{f}$ is the anomalous magnetic moment, given by
\beqa
\kappa_p & = & 1.79\left(\frac{|e|}{2m_p}\right)\,,\nonumber\\
\kappa_n & = & -1.91\left(\frac{|e|}{2m_n}\right)\,,
\eeqa
with $e$ being the electron charge. We have used the notation
$\sigma\cdot F = \sigma_{\mu\nu}F^{\mu\nu}$, where
$\sigma_{\mu\nu} = (i/2)[\gamma_\mu,\gamma_\nu]$, and $F^{\mu\nu}$
is the electromagnetic tensor for the $B$ field,
which we can write in the form
\beq
F_{\mu\nu} = iBP_{\mu\nu} \,,
\eeq
with
\beq
\label{defP}
P_{\mu\nu} = i\epsilon_{\mu\nu\alpha\beta}\,b^\alpha u^\beta \,.
\eeq
In an arbitrary frame of reference,
the vector $u^\mu$ in \Eq{defP} is the velocity four-vector of the medium.
In the frame in which the medium is at rest, which we adopt,
$u^\mu$ and $b^\mu$ take the form
\beqa
u^\mu = (1,\vec 0)\,,\nonumber\\
b^\mu = (0,\hat b)\,,
\eeqa
and the magnetic field is given by
\beq
\vec B = B\hat b\,.
\eeq
Furthermore, for definiteness, we choose the gauge such that
\beq
\label{gauge}
A_\mu = -\frac{1}{2}F_{\mu\nu}x^\nu \,.
\eeq

The idea now is to write
\beq
S(x,x^\prime) = \phi(x,x^\prime)\int\,\frac{d^4 p}{(2\pi)^4}
e^{-ip\cdot(x - x^\prime)} S_F(p) \,,
\eeq
where
\beq
\label{defphi}
\phi(x,x^\prime) = e^{\frac{i}{2}e_f x^\mu F_{\mu\nu}x^{\prime\nu}} \,,
\eeq
and substitute it in \Eq{coordeq} to obtain an equation for the
momentum space propagator $S_F(p)$. Using
\beq
\label{phieq}
i\partial_\mu\phi = -\frac{e_f}{2}F_{\mu\nu}x^{\prime\nu}\phi \,,
\eeq
it follows that
\beq
\label{covder}
(i\partial_\mu - e_fA_\mu)S(x,x^\prime) = \phi\int\,\frac{d^4 p}{(2\pi)^4}
e^{-ip\cdot(x - x^\prime)} \left[p_\mu - \frac{ie_f}{2}F_{\mu\nu}
\frac{\partial}{\partial p_\nu}\right]S_F(p) \,,
\eeq
and therefore this procedure leads to 
\beq
\label{Speq}
\left[\lslash{p} - \frac{ie_f}{2}F^{\mu\nu}
\gamma_\mu\frac{\partial}{\partial p^\nu} - m_f - 
\frac{1}{2}\kappa_f\sigma\cdot F\right]S_F(p) = 1 \,.
\eeq
This equation continues to hold in another gauge, different
from \Eq{gauge}, as long as $\phi$ is chosen such that is satisfies
\Eq{phieq}. In general, in another gauge $\phi$ is not 
be given by \Eq{defphi}.

Since we are interested in calculating the effects that are linear in $B$,
to solve \Eq{Speq} we substitute
\beq
\label{SFsplit}
S_F(p) = S_0 + S_B \,,
\eeq
where
\beq
\label{S0}
S_0 = \frac{1}{\lslash{p} - m_f + i\epsilon} \,,
\eeq
and then solve for $S_B$ by iteration. This yields
\beq
S_B = S_0\left(
\frac{ie_f}{2}F^{\mu\nu}\gamma_\mu\frac{\partial}{\partial p^\nu} +
\frac{1}{2}\kappa_f\sigma\cdot F\right)S_0 \,,
\eeq
which after using the relation
\beq
\frac{1}{2}\sigma\cdot F = B\gamma_5\lslash{u}\lslash{b} 
\eeq
and a little bit of Dirac algebra, can be written in the form
\beq
\label{SB}
S_B = \frac{1}{(p^2 - m^2_f + i\epsilon)^2}[e_f B G + \kappa_f B H] \,,
\eeq
where
\beqa
\label{GH}
G(p) & = & \gamma_5[(p\cdot b)\lslash{u} - (p\cdot u)\lslash{b} + 
m_f\lslash{u}\lslash{b}] \,,\nonumber\\[12pt]
H(p) & = & (\lslash{p} + m_f)\gamma_5\lslash{u}\lslash{b}(\lslash{p} + m_f) \,.
\eeqa

The thermal propagator, which incorporates the presence of the
nucleons in the background, is given as usual by\footnote{See, e.g., 
Ref.\ \cite{elmfors}.}
\beq
\label{Sfthermal}
S_f = S_F - [S_F - \bar S_F]\eta_f \,,
\eeq
where
\beq 
\label{etaf} 
\eta_f(p\cdot u) = \theta(p\cdot u)f_f(p\cdot u) +
\theta(-p\cdot u)f_{\bar f}(-p\cdot u)\,, 
\eeq
with
\beqa 
f_f(x) & = & \frac{1}{e^{\beta(x - \mu_f)} + 1}\,, \nonumber\\[12pt]
f_{\bar f}(x) & = & \frac{1}{e^{\beta(x + \mu_f)} + 1} \,. 
\eeqa
Here $\beta$ stands for the inverse temperature and $\mu_f$ for the
chemical potential of the nucleon $f$. 
Using Eq.\ (\ref{SFsplit}), together with
(\ref{SB}) and (\ref{GH}), the complete propagator, linear in $B$, 
can be expressed as
\beq
\label{Sfexpanded}
S_f = S_0 + S_B + S_T + S_{TB} \,,
\eeq
where $S_0$ and $S_B$ are given in \Eqs{S0}{SB}, while
\beqa
\label{STB}
iS_T & = & -2\pi\delta(p^2 - m^2_f)\eta_f (\lslash{p} + m_f)\nonumber\\
iS_{TB} & = & 2\pi\delta^\prime(p^2 - m^2_f)\eta_f 
[e_f BG + \kappa_f BH] \,,
\eeqa
with $G$ and $H$ given in \Eq{GH}. 
The notation $\delta^\prime$ denotes the derivative of the
delta function with respect to its argument. 

As a simple verification of these formulas,
we can consider the case of the electron, so that we can set
$\kappa_f = 0$ and $e_f = -|e|$ in \Eqs{SB}{STB}. In this case the
resulting formula for
the propagator precisely coincides with the formula
that was obtained in \Ref{nievesI} by expanding
directly the Schwinger formula, as it should be.
We have also used this propagator to compute
the nucleon contribution to the neutrino self-energy in a matter
background in the presence of a $B$ field, and we have verified that
the known result \cite{dn2} is reproduced.

%
%
\section{Nucleon contribution to the vertex function}
\label{sec:vertexfunction}

\begin{figure}
\begin{center}
%
%
\begin{picture}(220,120)(-110,-60)
\Photon(40,0)(80,0){-3}{6}
\Photon(-40,0)(0,0){3}{6}
\ArrowLine(-60,-40)(-40,0) 
\ArrowLine(-40,0)(-60,40)
\LongArrowArc(20,0)(20,270,460)
\LongArrowArc(20,0)(20,90,280)
\Text(-60,-50)[]{$\nu(k)$}
\Text(-60,50)[]{$\nu(k^\prime)$}
\Text(-20,-12)[]{$Z(q)$}
\Text(60,-12)[]{$\gamma(q)$}
\Text(20,-30)[t]{$f(p)$}
\Text(20,30)[b]{$f(p + q)$}
\end{picture}
\caption[]{One-loop diagram for the contribution of a background 
nucleon $f = n,p$ to the neutrino electromagnetic vertex.
\label{fig:diagram}}
\end{center}
\end{figure}
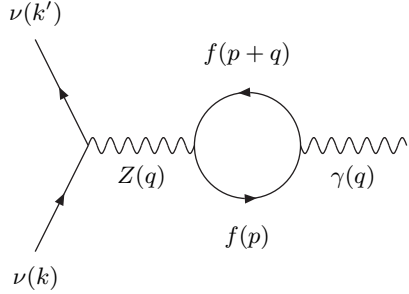

The total background-dependent part of the
vertex function for a neutrino of a given flavor $\ell = e,\mu\,\tau$
will be denoted by $\Gamma^{(\nu_\ell)}_\lambda$.
Here we denote by $\Gamma^{(e)}_\lambda$ the electron
contribution, which was calculated in \Ref{nievesI},
and write 
\beq
\Gamma^{(\nu_\ell)}_\lambda = \Gamma^{(e)}_\lambda + 
\Gamma^{(\mbox{nucl})}_\lambda \,.
\eeq
The subject of the present paper is the calculation of the 
nucleon contribution $\Gamma^{(\mbox{nucl})}_\lambda$, 
for which the relevant diagram is the one shown in Fig.\ \ref{fig:diagram}.

\subsection{Calculation of the vertex function}
\label{subsec:vertexfunctioncalc}
For each nucleon in the loop, the propagator is
given in \Eq{Sfexpanded}, while the electromagnetic couplings
are given by
\beq
L_{\gamma} = -|e|A^\mu \bar p\gamma_\mu p - 
\frac{\kappa_p}{2}\bar p\sigma^{\mu\nu}p F_{\mu\nu} - 
\frac{\kappa_n}{2}\bar n\sigma^{\mu\nu}n F_{\mu\nu} \,.
\eeq
For the neutral-current couplings we write
\beq
L_Z = - g_Z Z^\mu\left[
\sum_{\ell=e,\mu,\tau}\bar\nu_{L\ell}\gamma_\mu\nu_{L\ell} +
\sum_{f}\bar f\gamma_\mu(a_f + b_f\gamma_5)f\right] \,,
\eeq
where
\beqa
\label{Zcouplings}
g_Z & = & \frac{g}{2\cos\theta_W}\,,\nonumber\\
-a_e = a_p & = & \frac{1}{2} - 2\sin^2\theta_W\,, \nonumber\\
a_n & = & -\frac{1}{2}\,,\nonumber\\
b_e & = & \frac{1}{2}\,,\nonumber\\
b_n = -b_p & = & \frac{1}{2}g_A \,.
\eeqa
In \Eq{Zcouplings}, $g_A$
is the normalization constant of the axial-vector current of the nucleon,
$g_A = 1.26$. As depicted in the diagram, we denote by $k$ and $k^\prime$
the momentum of the incoming and outgoing neutrino, respectively,
and
\beq
q = k^\prime - k\,,
\eeq
is the momentum of the incoming photon.

With these conventions, the diagram in Fig.\ \ref{fig:diagram} 
leads to
\beq
\Gamma^{(\mbox{nucl})}_\lambda = -\frac{g^2}{4M^2_W}
(T^{(p)}_{\mu\lambda} + T^{(n)}_{\mu\lambda})\gamma^\mu L \,,
\eeq
where $L = \frac{1}{2}(1 - \gamma_5)$ as usual, and 
\beq
\label{Tfdef}
T^{(f)}_{\mu\lambda} = i\int\frac{d^4p}{(2\pi)^4}
\Tr\left[\gamma_\mu(a_f + b_f\gamma_5)
S_f(p + q)j^{(\mbox{em})}_{f\lambda}S_f(p)\right]\,.
\eeq
In \Eq{Tfdef}, $j^{(\mbox{em})}_{f\lambda}$ is the total electromagnetic
current of each nucleon,
\beqa
j^{(\mbox{em})}_{p\lambda} & = & |e|\gamma_\lambda + 
i\kappa_p\sigma_{\lambda\nu}q^\nu \,,\nonumber\\
j^{(\mbox{em})}_{n\lambda} & = & 
i\kappa_n\sigma_{\lambda\nu}q^\nu \,.
\eeqa

When \Eq{Sfexpanded} is substituted in \Eq{Tfdef} several terms are
produced. We are interested in the ones that depend on $B$ and
the distribution functions which, schematically, involve
the products $S_0 S_{TB}$ or $S_T S_B$. Denoting
the sum of all such terms by $T^{\prime\,(f)}_{\mu\lambda}$, 
then
\beqa
T^{\prime\,(f)}_{\mu\lambda} & = & i\int\frac{d^4p}{(2\pi)^4}
\Tr\left[
\gamma_\mu(a_f + b_f\gamma_5)
S_0(p + q)j^{(\mbox{em})}_{f\lambda}S_{TB}(p)\right.\nonumber\\ 
&&\mbox{} +
\gamma_\mu(a_f + b_f\gamma_5)
S_{TB}(p + q)j^{(\mbox{em})}_{f\lambda}S_{0}(p)\nonumber\\ 
&&\mbox{} +
\gamma_\mu(a_f + b_f\gamma_5)
S_{B}(p + q)j^{(\mbox{em})}_{f\lambda}S_{B}(p)\nonumber\\ 
&&\mbox{} +
\left.\gamma_\mu(a_f + b_f\gamma_5)
S_{B}(p + q)j^{(\mbox{em})}_{f\lambda}S_{B}(p)\right]\,.
\eeqa
Using the formulas given \Eqss{S0}{SB}{STB}, 
$T^{\prime\,(f)}_{\mu\lambda}$ can be written in the form
\beq
\label{Tfsplit}
T^{\prime\,(f)}_{\mu\lambda} = a_f T^{(fV)}_{\mu\lambda} - 
b_f T^{(fA)}_{\mu\lambda} \,,
\eeq
where
\beqa
\label{TVAdef}
T^{(fV)}_{\mu\lambda} & = & -4B \int\frac{d^4p}{(2\pi)^3} \eta_f(p.u)\left\{
\frac{-L^{(1)}_{\mu\lambda}\delta^\prime(p^2 - m^2_f)}{(p + q)^2 - m^2_f} +
\frac{L^{(2)}_{\mu\lambda}\delta(p^2 - m^2_f)}{[(p + q)^2 - m^2_f]^2}
- (q\rightarrow -q)\right\} \,,\nonumber\\
T^{(fA)}_{\mu\lambda} & = & -4B \int\frac{d^4p}{(2\pi)^3} \eta_f(p.u)\left\{
\frac{-K^{(1)}_{\mu\lambda}\delta^\prime(p^2 - m^2_f)}{(p + q)^2 - m^2_f} +
\frac{K^{(2)}_{\mu\lambda}\delta(p^2 - m^2_f)}{[(p + q)^2 - m^2_f]^2}
+ (q\rightarrow -q)\right\} \,,
\eeqa
with the the definitions
\beqa
\label{LKdef}
4L^{(1)}_{\mu\lambda}(p,q) & = & \Tr\gamma_\mu(\lslash{p} + \lslash{q} + m_f)
j^{\mbox{(em)}}_{f\lambda}[e_f G(p) + \kappa_f H(p)]\,,\nonumber\\
4L^{(2)}_{\mu\lambda}(p,q) & = & \Tr\gamma_\mu[e_f G(p) + \kappa_f H(p)]
j^{\mbox{(em)}}_{f\lambda}(\lslash{p} + \lslash{q} + m_f)\,,\nonumber\\
4K^{(1)}_{\mu\lambda}(p,q) & = & 
\Tr\gamma_5\gamma_\mu(\lslash{p} + \lslash{q} + m_f)
j^{\mbox{(em)}}_{f\lambda}[e_f G(p) + \kappa_f H(p)]\,,\nonumber\\
4K^{(2)}_{\mu\lambda}(p,q) & = & 
\Tr\gamma_5\gamma_\mu[e_f G(p) + \kappa_f H(p)]
j^{\mbox{(em)}}_{f\lambda}(\lslash{p} + \lslash{q} + m_f)\,.
\eeqa
In arriving at \Eq{TVAdef} we have applied the usual trick of making
the change of variable $p\rightarrow p - q$ in those terms
of the integrand that contain the factor $\eta_f((p + q)\cdot u)$,
and used the relations
\beqa
\label{LKrelations}
L^{(1)}_{\mu\lambda}(p - q,q) & = & -L^{(2)}_{\lambda\mu}(p,-q)\,,\nonumber\\
K^{(1)}_{\mu\lambda}(p - q,q) & = & K^{(2)}_{\lambda\mu}(p,-q)\,,
\eeqa
and the corresponding ones with the indices 1 and 2 interchanged.
The relationships given in \Eq{LKrelations}
can be established by manipulating the order of the factors
in the traces defined in \Eq{LKdef} with the help of 
the identity $C^{-1}\gamma_\mu C = -\gamma^T_\mu$, where
$C$ is the charge conjugation matrix ($C = i\gamma_2\gamma_0$ in the
Dirac representation).

\subsection{Structure of the vertex function}
\label{subsec:vertexfunctionstruct}
It follows from inspection of \Eq{TVAdef} that $T^{(fV)}_{\mu\lambda}$
transforms as a pseudo-tensor under parity. Further, 
$T^{(fV)}_{\mu\lambda}$ and $T^{(fA)}_{\mu\lambda}$
satisfy the transversality relations
\beqa
\label{transversalityrelations}
q^\lambda T^{(fV)}_{\mu\lambda} = q^\mu T^{(fV)}_{\mu\lambda} & = & 0\,,
\nonumber\\
q^\lambda T^{(fA)}_{\mu\lambda} & = & 0 \,,
\eeqa
which follow from the
conservation of the electromagnetic current and the vector neutral-current.
We do not show it here, but in fact they can
be proven explicitly in the present context by starting from \Eq{Tfdef} 
and then using the equation satisfied by the propagator, given in \Eq{Speq}.

The most efficient way to calculate $T^{(fV)}_{\mu\lambda}$ and 
$T^{(fA)}_{\mu\lambda}$ is to decompose them 
in terms of a complete set of tensors, consistently
with the above properties, and then evaluate the corresponding form factors. 
For this purpose it is useful to introduce the vectors
\beqa
\label{tildeubt}
\tilde u_\mu & = & \tilde g_{\mu\nu} u^\nu\,, \nonumber\\
\tilde b_\mu & = & \tilde g_{\mu\nu} b^\nu -\frac{q\cdot u}{q^2}
\left(q\cdot u \, b_\mu - q\cdot b \, u_\mu \right)\,,\\
\tilde t_\mu & = & 
\epsilon_{\mu\alpha\beta\gamma} u^\alpha b^\beta q^\gamma \,,\nonumber
\eeqa
where
\beq
\tilde g_{\mu\nu} \equiv g_{\mu\nu} - \frac{q_\mu q_\nu}{q^2}\,,
\eeq
which are orthogonal to $q^\mu$. Following \Ref{nievesI},
we can decompose $T^{(fV)}_{\mu\lambda}$ and $T^{(fA)}_{\mu\lambda}$
in the form
\beqa
\label{TVform}
T^{(fV)}_{\mu\lambda} & = & 
\tilde T^{(fV)}_1 \tilde P_{1\mu\lambda} + 
\tilde T^{(fV)}_2 \tilde P_{2\mu\lambda} + 
\tilde T^{(fV)}_S S_{2\mu\lambda}\,,\\
\label{TAform}
T^{(fA)}_{\mu\lambda} & = & 
T^{(fA)}_L Q_{\mu\lambda} + 
T^{(fA)}_T \left(\tilde g_{\mu\lambda} - Q_{\mu\lambda}\right) + 
T^{(fA)}_A A_{3\mu\lambda} +
T^{(fA)}_S S_{3\mu\lambda} + 
T^{(fA)}_u q_\mu \tilde u_\lambda + 
T^{(fA)}_b q_\mu \tilde b_\lambda \,,
\eeqa
where the various tensors that appear here are defined as follows,
\beqa
\label{tensors}
Q_{\mu\lambda} & = & 
\frac{\tilde u_\mu \tilde u_\lambda}{\tilde u^2}\,,\nonumber\\
S_{2\mu\lambda} & = & 
\tilde u_\mu \tilde t_\lambda + (\mu\leftrightarrow\lambda)\,,\nonumber\\
S_{3\mu\lambda} & = & 
\tilde u_\mu \tilde b_\lambda + (\mu\leftrightarrow\lambda) \,,\nonumber\\
A_{3\mu\lambda} & = & 
\tilde u_\mu \tilde b_\lambda - (\mu\leftrightarrow\lambda) \,,\nonumber\\
\tilde P_{1\mu\lambda} & = & i\epsilon_{\mu\lambda\alpha\beta}
\tilde u^\alpha q^\beta \,,\nonumber\\
\tilde P_{2\mu\lambda} & = & i\epsilon_{\mu\lambda\alpha\beta}
\tilde b^\alpha q^\beta \,.
\eeqa
In \Eq{TVform} we have included
a term proportional to $S_{2\mu\lambda}$ because, in contrast
to the case of the electron background contribution, 
$T^{(fV)}_{\mu\lambda}$ is not antisymmetric.

The form factors are functions of the
three scalar variables $\omega$, $Q_\parallel$ and $Q_\perp$ which
are defined by
\beqa
\label{kinematicvars}
\omega & = & q\cdot u \,, \nonumber\\
Q_\parallel & = & -q\cdot b\,,\nonumber\\
Q_\perp & = & \sqrt{Q^2 - Q^2_\parallel}\,,
\eeqa
where
\beq
Q^2 = |\vec Q|^2 = \omega^2 - q^2\,.
\eeq
The integral expressions for the form factors are obtained by contracting
both sides of \Eqs{TVform}{TAform} with the various tensors
that appear in those expansions. For example, the form factors
$T^{(fA)}_L$ and $T^{(fA)}_u$, which enter in the definition
of the neutrino induced charge, are given by
\beqa
\label{TALu}
T^{(fA)}_L & = & 
-\frac{4B}{\tilde u^2} \int\frac{d^4p}{(2\pi)^3} \eta_f(p.u)\left\{
\frac{-\tilde u^\mu\tilde u^\lambda K^{(1)}_{\mu\lambda}
\delta^\prime(p^2 - m^2_f)}{(p + q)^2 - m^2_f} +
\frac{\tilde u^\mu\tilde u^\lambda K^{(2)}_{\mu\lambda}
\delta(p^2 - m^2_f)}{[(p + q)^2 - m^2_f]^2} + (q\rightarrow -q)\right\}\,,
\nonumber\\
T^{(fA)}_u & = & 
-\frac{4B}{q^2 \tilde u^2} \int\frac{d^4p}{(2\pi)^3} \eta_f(p.u)\left\{
\frac{- q^\mu\tilde u^\lambda K^{(1)}_{\mu\lambda}
\delta^\prime(p^2 - m^2_f)}{(p + q)^2 - m^2_f} +
\frac{q^\mu\tilde u^\lambda K^{(2)}_{\mu\lambda}
\delta(p^2 - m^2_f)}{[(p + q)^2 - m^2_f]^2} - (q\rightarrow -q)\right\}\,.
\eeqa

In the general case,
the formulas so obtained are cumbersome and not very illuminating,
and for this reason we do not proceed any further along this direction.
Instead, we consider specifically the calculation of 
the form factors in the static and long wavelength limit, 
which are related to the effective neutrino charge in
the medium. In this way we are able to obtain approximate formulas 
that can be used in practical situations, and it will allow us to
compare the importance of the nucleon
background contribution, relative to the electron background.

%
%
\section{Neutrino effective charge}
\label{sec:neutrinocharge}

We consider the calculation of the induced charge of the neutrino, and
specifically the
parameter that was denoted by $e^\parallel_{\nu_\ell}$ in \Ref{nievesI}.
We denote by $\Gamma_\mu(\omega,Q_\perp,Q_\parallel)$ the neutrino
vertex function evaluated for an arbitrary value of $q$, and use a similar
notation for the form factors as well. With this notation,
$e^\parallel_{\nu_\ell}$ is defined by
\beq
e^\parallel_{\nu_\ell} = \frac{1}{2E_\nu}\Tr L\lslash{k}
\Gamma_0(0,0,Q_\parallel\rightarrow 0) \,,
\eeq
where $k^\mu = (E_\nu,\vec k)$ is the neutrino momentum vector. 

\subsection{Nucleon contribution}
Using the same notation for the nucleon contribution,
the quantity that we wish to calculate is
\beq
\label{enuclpar}
e^\parallel_{\mbox{nucl}} = \frac{1}{2E_\nu}\Tr L\lslash{k}
\Gamma^{\mbox{(nucl)}}_0(0,0,Q_\parallel\rightarrow 0) \,.
\eeq
In practice, we can evaluate the vertex function in this limit
by first setting
\beq
\label{trick}
q^\mu = Q_\parallel b^\mu \,,
\eeq
and then taking $Q_\parallel\rightarrow$ afterwards.
From the definition of the various tensors given in \Eq{tensors},
it then follows that
\beq
\label{Gammazeronucl}
\Gamma^{\mbox{(nucl)}}_0(0,0,Q_\parallel) = \frac{g^2}{4M^2_W}\sum_{f = n,p}
b_f\left\{T^{(fA)}_L\gamma_0 - 
Q_\parallel T^{(fA)}_u\vec\gamma\cdot\hat b\right\}L \,,
\eeq
with the form factors evaluated at $\omega = 0$ and $Q_\perp = 0$.
Therefore, to evaluate them, we can put $\tilde u_\mu \rightarrow u_\mu$
in \Eq{TALu}. We now use a property of the traces that are defined
in \Eq{LKdef}, which we can state precisely by defining
\beq
K^{(i)}_L(\vec p, \vec q) \equiv u^\mu u^\lambda K^{(i)}_{\mu\lambda} 
\qquad (i = 1,2)\,,
\eeq
where we have indicated explicitly the dependence on the spatial components
of the momentum vectors. Then, with $q^\mu$ as given in \Eq{trick}, 
it follows from the trace formula in \Eq{LKdef} that
\beq
\label{KsubLeq0}
K^{(i)}_L(-\vec p, -\vec q) = - K^{(i)}_L(\vec p, \vec q) \,.
\eeq
Using this property, and the fact that the distribution functions 
are isotropic in $\vec p$, \Eq{TALu} yields
\beq
\label{TALfinal}
T^{(fA)}_L(0,0,Q_\parallel) = 0 \,.
\eeq

Thus, we need to evaluate only $T^{(fA)}_u$, which
we write in the form
\beq
\label{TAuaux}
T^{(fA)}_u = -\frac{4B}{q^2\tilde u^2}\left[I_1 + I_2  - 
(q\rightarrow -q)\right]\,,
\eeq
where
\beqa
I_1 & = & (-1)\int\frac{d^4p}{(2\pi)^3} \eta_f(p.u)
\frac{F_1(p,q)\delta^\prime(p^2 - m^2_f)}{(p + q)^2 - m^2_f}\,,\nonumber\\
I_2 & = & \int\frac{d^4p}{(2\pi)^3} \eta_f(p.u)
\frac{F_2(p,q)\delta(p^2 - m^2_f)}{[(p + q)^2 - m^2_f]^2} \,,
\eeqa
with
\beq
\label{F12}
F_i(p,q) \equiv q^\mu \tilde u^\lambda K^{(i)}_{\mu\lambda}
\qquad (i = 1,2) \,.
\eeq
In order to calculate he traces that are involved
in \Eq{F12}, it is useful to note that if we calculate
\beq
F_1 = \frac{1}{4}q^\mu \tilde u^\lambda
\Tr\gamma_5\gamma_\mu(\lslash{p}^\prime + m_f)
j^{\mbox{(em)}}_{f\lambda}[e_f G(p) + \kappa_f H(p)]\,,
\eeq
where we have put $p^\prime = p + q$, then the result for $F_2$ is
obtained by making the substitution $p^\prime \leftrightarrow -p$.
After some tedious, but straightforward
Dirac algebra, in this way we obtain, for $q$ as given in \Eq{trick},
\beqa
\label{F12formulas}
F_1 & = & e_f^2 A + e_f \kappa_f B_1 + \kappa^2_f C_2 \,,\nonumber\\
F_2 & = & e_f^2 A + e_f \kappa_f B_2 + \kappa^2_f C_2 \,,
\eeqa
where
\beqa
\label{ABC}
A & = & -Q_\parallel\left(p^{0\,2} + P^2_\parallel + m^2_f +
Q_\parallel P_\parallel\right) \,,\nonumber\\
B_2 & = & m_f Q_\parallel (p^2 - m^2_f) -
2m_f Q_\parallel(2p^{0\,2} - Q^2_\parallel) \,,\nonumber\\
B_1 & = & B_2 - m_f Q^2_\parallel(2P_\parallel + Q_\parallel)\,,\nonumber\\
C_1 & = & Q^2_\parallel\left[-(3P_\parallel + Q_\parallel)(p^2 - m^2_f)
+ 2(2P_\parallel + Q_\parallel)(p^{0\,2} - P^2_\parallel) - 4m^2_fP_\parallel
\right]\,,\nonumber\\
C_2 & = & -Q^2_\parallel\left[\vphantom{P^2_\parallel}
-(3P_\parallel + 2Q_\parallel)(p^{2} - m^2_f)
+ (2P_\parallel + Q_\parallel)\left(2p^{0\,2} - 
P_\parallel (2P_\parallel + Q_\parallel)\right)
- 4m^2_f(P_\parallel + Q_\parallel)\right] \,,
\eeqa
with
\beq
P_\parallel = \vec p\cdot\hat b \,.
\eeq
To calculate $T^{(fA)}_u(0,0,Q_\parallel)$ 
we use the long wavelength limit expressions for $I_{1,2}$
that were derived in \Ref{nievesI}, which are
useful for computing the integrals in the limit that we need.
Those expressions for $I_{1,2}$ are obtained by using the
auxiliary formula
\beq
\label{I12lwaux}
\int\frac{d^3p}{(2\pi)^3}\frac{{\cal F}(p)}{[q^2 + \lambda 2p\cdot q]^n} = 
\lambda^n \int\frac{d^3p}{(2\pi)^3}\frac{\left({\cal F} - 
\lambda\frac{\vec Q}{2}\cdot\frac{d{\cal F}}{d\vec p} - \lambda
\frac{n\omega}{2E}{\cal F}\right)}
{[2E\omega - 2\vec p\cdot\vec Q]^n} \,,
\eeq
where $\lambda = \pm 1$, and they
are valid for $q\ll\langle{\cal E}\rangle$, 
where $\langle{\cal E}\rangle$ denotes a 
typical average energy of the particles in the background. 
Thus\footnote{We take the opportunity to point out that in
Ref.\ \cite{nievesI}, Eq. (5.1) and the second formula in Eq. (5.6)
are not written correctly. The correct formulas are given here in
\Eqs{I12lwaux}{I12lw}, respectively. In the calculations of 
Ref.\ \cite{nievesI}, the correct formulas were used.},
\beqa
\label{I12lw}
I_1 & = & \int\frac{d^3p}{(2\pi)^3}\left\{
\frac{F^{\prime -}_1 - \frac{\vec Q}{2}\cdot\frac{d F^{\prime +}_1}{d\vec p}
- \frac{\omega}{2E}F^{\prime +}_1}{2E\omega - 2\vec p\cdot\vec Q} -
\frac{F^{+}_1 - \frac{\vec Q}{2}\cdot\frac{d F^{-}_1}{d\vec p}}
{[2E\omega - 2\vec p\cdot\vec Q]^2}\right\} \,,\nonumber\\
I_2 & = & \int\frac{d^3p}{(2\pi)^3}
\frac{F^{+}_2 - \frac{\vec Q}{2}\cdot\frac{d F^{-}_2}{d\vec p}
- \frac{\omega}{E}F^{-}_2}{[2E\omega - 2\vec p\cdot\vec Q]^2} \,,
\eeqa
where
\beqa
\label{F12pm}
F^{\pm}_1(p,q) & = & \left.\left[\frac{f_e(p^0)F_1(p,q) \pm
f_{\bar e}(p^0)F_1(-p,q)}{2p^0}\right]\right|_{p^0 = E(p)}\,, \nonumber\\
F^{\pm}_2(p,q) & = & \left.\left[\frac{f_e(p^0)F_2(p,q) \pm
f_{\bar e}(p^0)F_2(-p,q)}{2p^0}\right]\right|_{p^0 = E(|\vec p|)}\,,\nonumber\\
F^{\prime\pm}_1(p,q) & = & \left.\left\{\frac{1}{2p^0}
\frac{\partial}{\partial p^0}\left[
\frac{f_e(p^0)F_1(p,q) \pm f_{\bar e}(p^0)F_1(-p,q)}{2p^0}
\right]\right\}\right|_{p^0 = E(|\vec p|)}\,.
\eeqa
In \Eq{I12lw} the symbol $\frac{d}{d\vec p}$ stands for
the total momentum derivative,
\beq
\frac{d}{d\vec p} = \frac{\partial}{\partial\vec p} + \frac{\vec p}{E}
\frac{\partial}{\partial E} \,.
\eeq
The final step consists in putting 
$\omega = 0$ and $\vec Q = Q_\parallel\hat b$ in \Eq{I12lw}
and using \Eqs{F12formulas}{ABC} to determine the various terms
in the integrands for $I_{1,2}$. The details of this calculation are similar
to the corresponding ones for the case of the electron background,
some of which were given in \Ref{nievesI}, and which we omit here.
In this way we arrive at
\beq
I_1 + I_2 - (q \rightarrow -q) = -Q_\parallel\left[
e^2_f t^{(f)}_\parallel + e_f \kappa_f (2m_f t^{(f)}_\parallel) +
\kappa^2_f s^{(f)}_\parallel\right] + O(Q^3_\parallel) \,,
\eeq
where we define, for any fermion $f$,
\beqa
\label{tf}
t^{(f)}_\parallel & = & -\frac{1}{2}\int^\infty_0\frac{dp}{(2\pi)^2}
\frac{\partial}{\partial E}\left[f_f(E) + f_{\bar f}(E)\right]\,,\\[12pt]
\label{sf}
s^{(f)}_\parallel & = & \int\frac{d^3p}{(2\pi)^3}
\frac{f_f(E) + f_{\bar f}(E)}{E}\,.
\eeqa
Therefore, from \Eq{TAuaux},
\beq
\label{TAufinal}
T^{(fA)}(0,0,Q_\parallel\rightarrow 0) = -\frac{4B}{Q_\parallel}
\left[e_f(e_f + 2m_f\kappa_f)t^{(f)}_\parallel + 
\kappa^2_f s^{(f)}_\parallel\right]\,,
\eeq
and finally, from \Eqs{enuclpar}{Gammazeronucl}, 
and using the results given in \Eqs{TALfinal}{TAufinal},
we obtain
\beq
\label{enuclfinal}
e^\parallel_{\mbox{nucl}} = \frac{g^2}{M^2_W}(\vec B\cdot\hat k)
\left[b_n \kappa^2_n s^{(n)}_\parallel + 
b_p (|e|^2 + 2m_f|e|\kappa_p)t^{(p)}_\parallel
+ b_p \kappa^2_p s^{(p)}_\parallel\right] \,.
\eeq

\subsection{Discussion}
In the environments of physical interest the nucleons are
non-relativistic. Therefore, in \Eq{sf} we can effectively
replace $E\rightarrow m_f$ in the denominator, so that
for the nucleons
\beq
s^{(f)}_\parallel = \frac{n_{f}}{2m_f}\,,
\eeq
where $n_f$ is the total number density of the neutrons
or protons. 
Using the result obtained in \Ref{nievesI} for the electron background term,
the total matter contribution to the
$B$-dependent part the induced neutrino charge is then
\beq
\label{eBtotal}
e^\parallel_{\nu_\ell} = \frac{g^2}{M^2_W}(\vec B\cdot\hat k)\left[
-e^2 \chi^{(\ell)}_A t^{(e)}_\parallel +
\frac{1}{2m_n}b_n \kappa^2_n n_n + \frac{1}{2m_p}b_p \kappa^2_p n_p + 
b_p (|e|^2 + 2m_p|e|\kappa_p)t^{(p)}_\parallel
\right]\,,
\eeq
with $t^{(e,p)}_\parallel$ as defined in \Eq{tf}, and
\beq
\chi^{(\ell)}_A = \delta_{\ell,e} - b_e \,.
\eeq
The additional term that is present for the electron neutrino ($\ell = e$)
is due to the charged-current interaction, which is absent for 
$\nu_{\mu,\tau}$.

The evaluation of $t^{(e,p)}_\parallel$
requires more care, and we must distinguish various cases.
For a relativistic gas, putting $E\simeq p$ in \Eq{tf} yields
\beq
\label{tfrel}
t^{(f)}_\parallel = \frac{1}{8\pi^2} \qquad (\mbox{relativistic})\,.
\eeq
For a non-relativistic (NR) gas, we distinguish two cases. If the gas
can be treated classically, then using the relation
\beq
\frac{\partial f}{\partial E} = -\beta f \,,
\eeq
yields
\beq
\label{tfnrclass}
t^{(f)}_\parallel = \frac{n_f\beta^2}{8 m_f} \qquad (\mbox{NR classical)} \,.
\eeq
On the other hand, if the gas is degenerate, we obtain the approximate
formula
\beq
\label{tfnrdeg}
t^{(f)}_\parallel = \frac{m_f}{8\pi^2 p_F}\left[1 + \frac{\pi^2}{8}
\frac{m^2_f}{\beta^2 p^4_F}\right] \qquad (\mbox{NR degenerate})\,,
\eeq
where $p_F$ is the Fermi momentum, for
\beq
p_F \gg \sqrt{m_f/\beta} \,.
\eeq
The derivation of \Eq{tfnrdeg} is sketched in the Appendix.

The relative importance of the various terms in \Eq{eBtotal} depends
on the value of $t^{(e)}_\parallel$ and $t^{(p)}_\parallel$
which in turn depend on the conditions of the environment. 
The above approximate formulas are useful in this respect. 
For example, suppose that the conditions are such that both the electron and
proton gases can be treated classically. Then for the protons
we use \Eq{tfnrclass}. If also the electrons are non-relativistic
then we use the same formula for them and,
remembering that the charge neutrality of the medium
requires that $n_p = n_e$, it follows that
\beq
\frac{t^{(p)}_\parallel}{t^{(e)}_\parallel} = \frac{m_e}{m_p} \,.
\eeq
On the other hand, if the electrons are relativistic, 
using \Eq{tfrel} for them yields instead
\beq
\frac{t^{(p)}_\parallel}{t^{(e)}_\parallel} \sim \frac{T}{m_p} \,.
\eeq
In either case, we can neglect the $t^{(p)}_\parallel$ term in \Eq{eBtotal}.

In contrast, consider the case in which the gases are degenerate,
and for definiteness suppose that the electrons are relativistic.
Then using \Eq{tfnrdeg} for the protons and \Eq{tfrel} for
the electrons, it follows that $t^{(p)}_\parallel$ is
larger than $t^{(e)}_\parallel$ by a factor $m_p/p_{Fp}$.
Furthermore, the neutron term in \Eq{eBtotal}, as well as the
proton term that is proportional to $n_p$, are of order
$n_n/m^3_n \sim p^3_{Fn}/(3\pi^2 m^3_n)$ which,
for $p_{Fn}/m_n \sim 1/3$, is about a factor of 10 smaller than
the electron term $t^{(e)}_\parallel = 1/8\pi^2$.
Therefore, in this case the proton term may be the dominant one
in \Eq{eBtotal}.

In any application to the calculation of neutrino processes rates,
besides the $B$-dependent coupling just determined, the amplitude
contains a $B$-independent term. To have an idea of their relative
magnitude we can compare the value of $e^\parallel_\nu$ with the
analogous quantity $e^{(0)}_\nu$ calculated in the absence
of the $B$ field. From Ref.\ \cite{np2}, with a slight change
of notation,
\beq
\label{effchargenoB}
e^{(0)}_\nu = - \frac{g^2\chi^{(\ell)}_V}{4M^2_W}\frac{1}{|e|r^2_D}\,,
\eeq
where $\chi^{(\ell)}_V = 2\sim^2\theta_W - \frac{1}{2} + \delta_{\ell,e}$,
and the parameter $r_D$ is the Debye screening length, which is given
in terms of the longitudinal component of the photon self-energy
$\pi_L(\omega,Q)$ by
\beq
r^{-2}_D = \pi_L(0,Q\rightarrow 0) \,.
\eeq
Considering again the case in which the gases are degenerate, with the
electrons being relativistic, then we can take\footnote{See, e.g., 
Ref.\ \cite{raffeltbook}, p. 220.}
\beq
r^{-2}_D = \frac{e^2 m_p p_{Fp}}{\pi^2} \,,
\eeq
neglecting the electron contribution, which is smaller by a factor
of $p_{Fp}/m_p$. Writing
\beq
\frac{e^\parallel_{\nu_\ell}}{e^{(0)}_{\nu_\ell}} =
R_n + R_P + R_e \,,
\eeq
the relative size of the various contributions can be estimated 
using \Eqs{eBtotal}{effchargenoB}. For example, for the protons we find
\beq
R_p \sim \frac{m^2_e B}{p^2_{Fe}B_c}\,,
\eeq
where we have used $p_{FP} = p_{Fe}$ and
$B_c = m^2_e/|e| = 4\times 10^{13} G$, while for the
electrons it is a factor of $p_{Fe}/m_p$ smaller,
and it is even smaller for the neutrons as we mentioned above.
Since our calculations are based on a weak-field approximation,
the implicit assumption is that the magnetic contribution
$e^\parallel_{\nu_\ell}$ is smaller than the zeroth order term
$e^{(0)}_{\nu_\ell}$. Nevertheless, it is conceivable that
there are physical systems for which $e^\parallel_{\nu_\ell}$
is not negligible, which can produce observable effects due to its
anisotropic nature. The above discussion shows that, in those
situations, the nucleon couplings to the $B$ field must be
taken into account.
 %
%
\section{Conclusions}
\label{sec:conclusions}
A neutrino that propagates through a matter medium acquires an
induced electric charge due to its interactions with the matter
particles.
We have considered the $B$-dependent part of the neutrino electromagnetic
vertex, and more specifically the neutrino induced charge, in a background
of electrons and nucleons in the presence of a magnetic field.
We have extended the previous calculations,
which considered only the electrons in the background, 
to take into account the nucleons
and, in particular, to include the effects of their anomalous magnetic
coupling. The importance of the contribution due to the
nucleons, relative to the electron contribution,
depends on the conditions of the environment considered. For example,
we indicated that
in some situations the term due to the protons is not significant,
but there are cases in which the terms due to the nucleons are 
the most important ones.

Besides the term that we have calculated, which is linear in $B$,
the induced charge contains a term that is independent of $B$.
The distinctive feature of the $B$-dependent term is that
it is not isotropic in the neutrino momentum; it has a different
value for neutrinos going in different directions, relative to $\vec B$.
This could produce observable effects in the context of a variety of physical
problems that have been considered in the literature, such as those
related to the influence of the neutrino electromagnetic interactions
on the collective properties of a plasma.
Our results can be useful in those contexts, 
and this work paves the way
for considering and studying those effects in more detail.

\acknowledgments
This material is based upon work supported  by the US National
Science Foundation under Grant No. 0139538.

\appendix
%
%
\section{Proof of \Eq{tfnrdeg}}

Naively, in the completely degenerate limit we can put
\beq
\label{fermideg}
f_f = \Theta(E_F - E)
\eeq
in \Eq{tf} and, assuming that $f_{\bar f} \simeq 0$, obtain
\beq
t^{(f)}_\parallel = \frac{E_F}{8\pi^2 p_F} \,.
\eeq
This agrees with \Eq{tfrel} in the relativistic limit, and with the leading
term of \Eq{tfnrdeg} in the non-relativistic one,  
but this procedure leaves unanswered the question of the
range of validity of this approximation. For that, 
we consider in more detail the integral involving $f_f$ in \Eq{tf}. Similar
considerations apply to the other integral, involving the anti-particle
distribution, with obvious modifications. 

Defining
\beqa
\epsilon & = & E - m_f\,,\nonumber\\
\mu^\prime & = & \mu - m_f\,,
\eeqa
we write
\beqa
\label{Istart}
-\int^\infty_0 dp\, \frac{\partial}{\partial E}\frac{1}{e^{\beta(E - \mu)} + 1}
& = & \mbox{} -\int^\infty_0 dp\, \frac{\partial}{\partial\epsilon}
\frac{1}{e^{\beta(\epsilon - \mu^\prime)} + 1}\nonumber\\
& = & \frac{\partial}{\partial \mu^\prime} I\,,
\eeqa
where
\beq
\label{Idef}
I = \int^\infty_0 dp \frac{1}{e^{\beta(\epsilon - \mu^\prime)} + 1} \,,
\eeq
and changing the integration variable,
\beq
I = \int^\infty_0 d\epsilon 
\frac{g(\epsilon)}{e^{\beta(\epsilon - \mu^\prime)} + 1} \,,
\eeq
with
\beq
g(\epsilon) = \frac{E}{p} = 
\frac{\epsilon + m_f}{[(\epsilon + m_f)^2 - m^2_f]^{1/2}}\,.
\eeq

Letting
\beq
z = \beta(\epsilon - \mu^\prime) \,,
\eeq
$I$ can be written in the form
\beq
\label{I1}
I = \frac{1}{\beta}\int^0_{-\beta\mu^\prime}dz \,
g(\mu^\prime + \frac{z}{\beta}) f(z)
+ \frac{1}{\beta}\int^\infty_0 dz\, g(\mu^\prime + \frac{z}{\beta})f(z)
\eeq
where we have defined
\beq
f(z) = \frac{1}{e^z + 1} \,.
\eeq
In the first integral in \Eq{I1} we change $z\rightarrow -z$ and use
$f(-z) = 1 - f(z)$, which yields
\beq
I = \int^{\mu^\prime}_0 d\epsilon\,g(\epsilon) - 
\frac{1}{\beta}\int^{\beta\mu^\prime}_0 dz\, 
g(\mu^\prime - \frac{z}{\beta})f(z) +
\frac{1}{\beta}\int^\infty_0 dz\, g(\mu^\prime + \frac{z}{\beta})f(z)\,.
\eeq
Up to this point this expression is exact. The approximation now
consists in replacing the upper limit integration in the second integral
by infinity, and it is valid for 
\beq
\label{degcond}
\beta\mu^\prime \gg 1.
\eeq
This yields the final formula 
\beq
\label{Ifinal}
\frac{\partial}{\partial\mu^\prime}I = g(\mu^\prime) +
\frac{1}{\beta}\int^\infty_0 dz\, 
\frac{g^\prime(\mu^\prime + \frac{z}{\beta}) - 
g^\prime(\mu^\prime - \frac{z}{\beta})}{e^z + 1}\,,
\eeq
which can be evaluated explicitly by expanding $g(z)$ in a Taylor
series in $z$ and integrating term by term.

Apart from the condition given in \Eq{degcond}, there are no further
restrictions, so that this formula can be applied regardless
of whether the gas is relativistic or non-relativistic.
Thus, for example, in the non-relativistic limit,
$g(\epsilon) \approx (m_f/2\epsilon)^{1/2}$, and from \Eq{Ifinal}
\beq
\frac{\partial}{\partial\mu^\prime}I = \sqrt{\frac{m_f}{2\mu^\prime}}\left[
1 + \frac{\pi^2}{8\beta^2\mu^{\prime\,2}}\right] \,.
\eeq
Using this result in \Eq{Istart}, and substituting 
$\mu^\prime = p^2_F/2m_f$, leads to the formula quoted in \Eq{tfnrdeg}.

\end{document}